\documentclass[aps,prl,twocolumn,superscriptaddress,showpacs]{revtex4}

\usepackage{graphicx}
\usepackage{afterpage}

\begin{document}


\title{Ferromagnetism makes a doped semiconductor less shiny}


\author{F.P. Mena}
\affiliation{Materials Science Centre, University of Groningen,
9747 AG Groningen, The Netherlands}

\author{J.F. DiTusa}
\affiliation{Department of Physics and Astronomy, Louisiana State
University, Baton Rouge, Louisiana 70803 USA}

\author{D. van der Marel}
\affiliation{Materials Science Centre, University of Groningen,
9747 AG Groningen, The Netherlands} \affiliation{Departement de
Physique de la Matiere Condensee, Universite de Geneve, CH-1211
Geneve 4, Switzerland}

\author{G. Aeppli}
\affiliation{London Centre for Nanotechnology and Department of
Physics and Astronomy, UCL London WC1E6BT, UK}

\author{D.P. Young}
\affiliation{Department of Physics and Astronomy, Louisiana State
University, Baton Rouge, Louisiana 70803 USA}

\author{C. Presura}
\affiliation{Materials Science Centre, University of Groningen,
9747 AG Groningen, The Netherlands}

\author{A. Damascelli}
\affiliation{Department of Physics \& Astronomy, University of
British Columbia, Vancouver, B.C. V6T 1Z1, Canada}

\author{J.A. Mydosh}
\affiliation{Kamerling Onnes Laboratory, Leiden University, 2500
RA Leiden, The Netherlands,and Max-Plank-Institute for Chemical
Physics of Solids, D-01187 Dresden, Germany}


\date{\today}

\begin{abstract}
Magnetic semiconductors have attracted interest because of the
question of how a magnetic metal can be derived from a
paramagnetic insulator. Here our approach is to carrier dope
insulating FeSi and we show that the magnetic half-metal which
emerges has unprecedented optical properties, unlike those of
other low carrier density magnetic metals. All traces of the
semiconducting gap of FeSi are obliterated and the material is
unique in being less reflective in the ferromagnetic than in the
paramagnetic state, corresponding to larger rather than smaller
electron scattering in the ordered phase. \end{abstract}


\maketitle


Future technologies based on the control and state of electron
spins rather than charges are commonly referred to as spintronics.
Efforts to produce materials for spintronics have mostly focused
on thin film III-V semiconductors alloyed with manganese
\cite{mnd35}. In (GaMn)As, the most fully characterized of these
alloys, Mn substitutes a trivalent Ga ion and acts as a shallow
acceptor just above the valence band. The Mn$^{2+}$ impurities
have a local moment associated with a high spin (S=5/2)
configuration, and are ferromagnetically coupled below the Curie
temperature ($T_C$) by a small number of itinerant hole carriers.
In metallic and ferromagnetic (FM) (GaMn)As these doped holes
reside in an itinerant Mn-derived impurity band \cite{gamnasop}
about 0.1 eV above the valence band.

Another route to magnetic semiconductors relies on carrier doping
into small gap, strongly correlated insulators. The most
celebrated is the monosilicide FeSi, which has been investigated
for several decades because it has a large 300 K response to
magnetic fields that vanishes as $T$ approaches zero
\cite{manyala1,wernick,aepplirev}. Together with CoSi and the
unusual metal MnSi \cite{mena}, FeSi belongs to the larger group
of transition metal monosilicides, allowing chemical substitutions
across the entire series without change in the cubic B-20 crystal
structure or the nucleation of second phases
\cite{manyala1,wernick}. Bulk single crystals can be grown and
FeSi can be made metallic and FM, with electrical properties which
are unusually sensitive to external magnetic fields, by the
substitution of Co to form the silicon-based magnetic
half metal \cite{manyala1,guevera,menaprep} Fe$_{1-y}$Co${_y}$Si, 
whose calculated density of states is shown in Fig. 1.

Given that optical properties are both of fundamental interest and 
a key to potential spintronic applications \cite{wolf}, we have 
measured the optical conductivity $\sigma(\omega)$ of 
Fe$_{1-y}$Co$_y$Si. For clean semiconductors, the low-$T$ $\sigma(\omega)$ 
is dominated by excitations across a gap ($E_g$) between valence and conduction bands. 
Chemical doping adds carriers to the valence, conduction, 
or impurity bands and yields a $T=0$ $\sigma(\omega)$ which is very similar to 
that obtained by warming in the undoped case-–-the dominant low-$\omega$ feature is a 
Drude peak with weight proportional to the carrier density ($n$) and width $\Gamma$ 
measuring the typical scattering rate for the carriers. If the carriers were 
magnetically polarized their band would be 
split into majority and minority spin-bands. Eventually the minority band can 
move above the Fermi energy ($E_F$) resulting in a redistribution of all 
the carriers into the majority band. The outcome is a half metal, 
of which Fe$_{1-y}$Co${_y}$Si is an excellent example (Fig.\ 1), 
because electrons carrying the majority spin belong to a partially filled, 
metallic band, while those carrying the minority spin belong to an empty (at $T = 0$) 
conduction band \cite{degroot}. To accommodate all of the itinerant electrons $E_F$ 
shifts upward, decreasing the density of states (DOS) at $E_F$ by a factor of $2^{2/3}$. 
Naively one might expect this reduction in the DOS to reduce the low-$\omega$ $\sigma(\omega)$. 
However, the optical sum rule states that the integral of $\sigma(\omega)$ over $\omega$
measures $n$, so that conservation of $n$ in the parabolic band leads to 
conservation of $\sigma(\omega)$ through the phase transition as long as $\Gamma$ 
and the effective mass ($m^*$) of the carriers do not change. This is precisely 
what occurs for spin polarized metals such as CrO$_2$ and NiMnSb \cite{croop,nbmnsbop}. 
Similarly, data for (GaMn)As 
reveal changes to $\sigma(\omega)$ below $T_C$ due primarily to a reduction of $\Gamma$ and 
an associated reduction in $m^*$ \cite{gamnasop}. As a result $\sigma(\omega)$ 
increases below 3000 cm$^{-1}$ making the appearance more reflective. 
In addition $\sigma(\omega)$ of (GaMn)As displays a resonance at 2000 cm$^{-1}$ due to 
the promotion of electrons from the GaAs valence band to Mn acceptor levels which 
have been broadened into an impurity band \cite{gamnasop}.  Finaly, EuB$_6$ and 
La$_{1-x}$Sr$_x$MnO$_3$ display the most 
impressive increases in reflectivity, due to simultaneous ferromagnetic and 
insulator- to-metal transitions \cite{degiorgi,okimoto}.

\begin{figure}[tb]
 \includegraphics[angle=0,width=7cm, trim=0 20 0 0,clip=1]{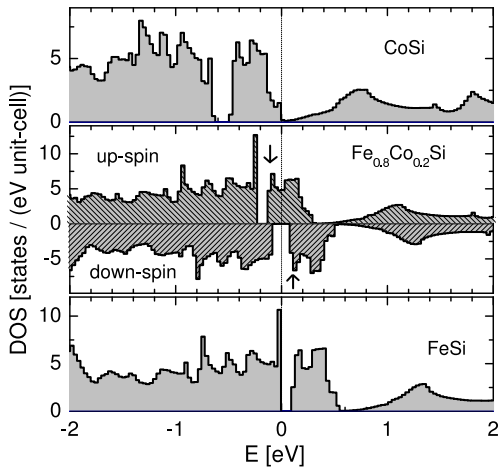}%
 \caption{\label{fig:dos} Calculated DOS. Electronic density of states 
calculated using the local density approximation (LDA) with the
Gunnarsson-Lundquist exchange correlation potential carried out
self consistently using the full potential linear muffin-tin
orbital method  \protect{\cite{menaprep}}. For Fe$_{0.8}$Co$_{0.2}$Si a
non-integer charge to iron in FeSi was assigned. 
Arrows are approximate $E_F$ in 
unpolarized state. }
 \end{figure}

We used vapor transport, light
image furnace floating zone, and modified tri-arc Czochralski
methods to grow single crystals from high purity starting materials.
X-ray spectra showed all samples to be single phase with
a lattice constant linearly dependent on $y$ demonstrating that Co
successfully replaces Fe over the entire concentration range
($0\le y\le 1$). Energy dispersive X-ray microanalysis yielded
results consistent with the nominal concentrations. 
Reflectivity ($R(\omega)$) was measured
from 30 to 6000 cm$^{-1}$ while ellipsometry gave
the dielectric function from 6000 to 36000 cm$^{-1}$.
Between 30 and 6000 cm$^{-1}$ we used Kramers-Kronig (KK)
relations, along with a Hagen-Rubens extrapolation of $R(\omega)$
data to $\omega = 0$, to obtain the phase of $R(\omega)$, and
subsequently $\epsilon(\omega)$. $\sigma(\omega)$ was found via
Re$\sigma = (\omega / 4\pi)$Im$\epsilon(\omega)$. We
have carefully checked that $\sigma(\omega)$ is not significantly
altered by our choice of high- and low-$\omega$ terminations. The
KK output was locked to the ellipsometrically measured
$\epsilon_1(\omega)$ and $\epsilon_2(\omega)$. 

Figs.\ 2 and 3 display $R(\omega)$ and $\sigma(\omega)$ of three
crystals for several $T$'s over a wide $\omega$ range. 
We begin
with pure CoSi (top frames in Figs.\ 2 and 3), long known as
a diamagnetic metal with a very low carrier density, $\sim 1\%$ of
electrons/formula unit \cite{asanabe}. Our data agree with the
simple ideas of $\sigma(\omega)$ of a low $n$ metal---there 
is a small Drude peak
centered at $\omega = 0$ which coexists with interband transitions
beginning at $\sim 1000$ cm$^{-1}$. At 10 K $\Gamma$= 165 cm$^{-1}$, which in the simplest
analysis implies a carrier mean free path of 5 nm. The main effect
of warming to 300 K is to raise $\Gamma$ by less
than $k_BT$.

\begin{figure}[tb]
 \includegraphics[angle=0,width=8cm]{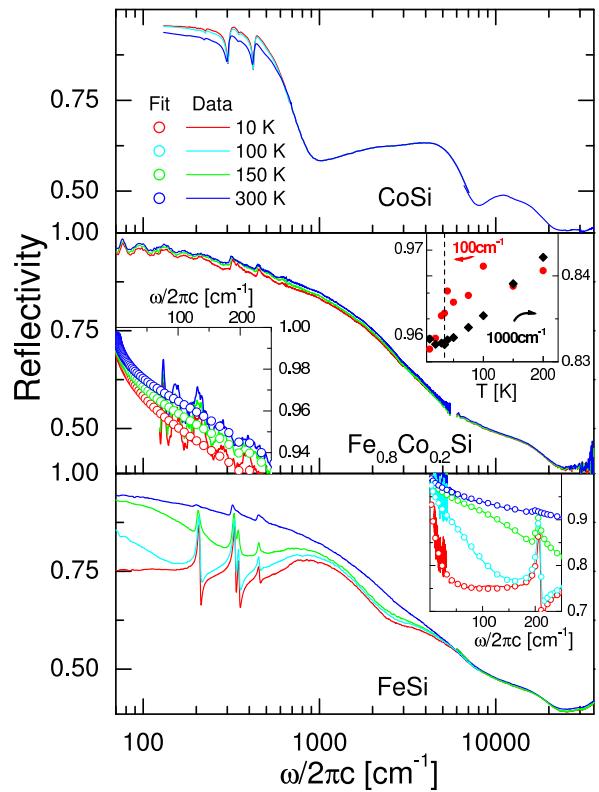}%
 \caption{\label{fig:reflectivities}
Near normal $R(\omega)$ of FeSi, CoSi, and Fe$_{0.8}$Co$_{0.2}$Si.
Insets: Expanded view of $R(\omega)$ (solid lines) and fit to
$R(\omega)$ at low-$\omega$ (open circles). Middle-panel inset:
$T$-dependence of $R(\omega)$ at 100 and 1000 cm$^{-1}$; dashed
line is $T_C$.}
 \end{figure}

 \begin{figure}[tb]
 \includegraphics[angle=0,width=8cm]{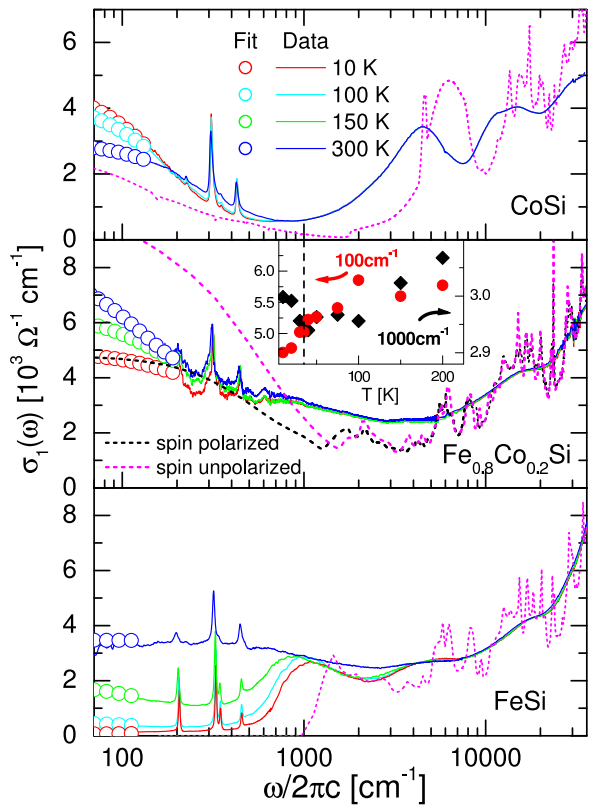}%
 \caption{\label{fig:sigma}
Real part of $\sigma(\omega)$ of FeSi, CoSi, and
Fe$_{0.8}$Co$_{0.2}$Si. (solid lines and open circles).
Dashed lines are $\sigma_1(\omega)$ obtained from the 
LDA band structure calculations,evaluated
\protect{\cite{menaprep}} with a $k$-space integration over
216 points in the irreducible part of the Brillouin zone.
The calculations fix both the positions of the interband transitions and
the weight of Drude peak. Widths ($\Gamma$) of the Drude
peaks were chosen to minimize deviation from experiment. For
Fe$_{0.8}$Co$_{0.2}$Si, polarized calculation is a much better
estimate for Drude weight than unpolarized calculation, even at
300 K, consistent with short range magnetic order well above
$T_C$. Notwithstanding the huge $\Gamma = 600$ cm$^{-1}$ used,
substantial spectral weight associated with Co doping is
unaccounted for. Inset: $T$-dependence of $\sigma_1(\omega)$ at
100 and 1000 cm$^{-1}$; dashed line is $T_C$. }
 \end{figure}

The lower frames of Figs.\ 2 and 3
display data for the alloy's other end member, insulating FeSi.
$\sigma(\omega)$ which shows no hint of a band gap at 300 K, is
almost completely suppressed below 600 cm$^{-1}$ upon cooling to
10 K, as in Ref. \cite{schlesinger} which emphasized that in a
conventional band picture, remnants of a 60 meV gap should be
clearly visible at 300 K (26 meV) \cite{schlesinger,damascelli}. A
second important feature is that the energy range over which
$\sigma(\omega)$ changes as $T$ is reduced is extremely large.
Estimates based on the data of Fig. \ref{fig:sigma} agree with
previous reports of a lack of spectral weight conservation to
energies above 80 times $E_g$ \cite{schlesinger}. These two
observations represent a miserable failure of the independent
electron model for FeSi \cite{urasaki}.

Doping FeSi via substitution of Co for Fe yields similar problems
for the standard model underlying semiconductor optics. In
contrast to the data for pure CoSi, but in agreement with our
findings for FeSi at 300 K, Fe$_{0.8}$Co$_{0.2}$Si displays a
$\sigma(\omega)$ which decays weakly from $\omega = 0$ to 3000
cm$^{-1}$, and all traces of the gap in the pure FeSi parent are
obliterated \cite{chernikov}. Even with the assumption of a
scattering rate in excess of the 60 meV gap of FeSi, a simple
Drude analysis (dashed lines in Fig.\ref{fig:sigma}) based on our
band structure calculations cannot account for $\sigma(\omega)$.
We conclude that treating the electrons in Fe$_{1-y}$Co$_y$Si as a
simple Fermi liquid formed in the conduction band is incorrect,
notwithstanding the remarkable simplicity of aggregate $\omega =
0$ properties such as the normal Hall effect and ordered
magnetization, which correspond to one carrier and one polarized
spin per Co atom \cite{manyala1}.

Beyond showing that the parent insulator and its electron-doped
derivative violate standard ideas about undoped and doped
semiconductors, Figs.\ 2 and 3 also reveal that
Fe$_{1-y}$Co$_y$Si defies expectations for itinerant magnets. In
particular, cooling yields a loss of spectral weight of a
different qualitative nature than seen for FeSi, where it occurs
throughout the gapped region. The loss is apparent not only in
$\sigma(\omega)$ derived via Kramers-Kronig from the raw data, but
also in the directly observed reflectivity. Thus, in contrast to
what occurs for all other metallic ferromagnets, including isostructural
MnSi \cite{mena} and (Ga,Mn)As \cite{gamnasop}, the approach and onset
of magnetic order at $T_C = 36$ K decreases the reflectivity (shininess)
of  Fe$_{1-y}$Co$_y$Si.

 \begin{figure}[tb]
 \includegraphics[angle=0,width=7cm, trim=0 60 0 0,clip=1]{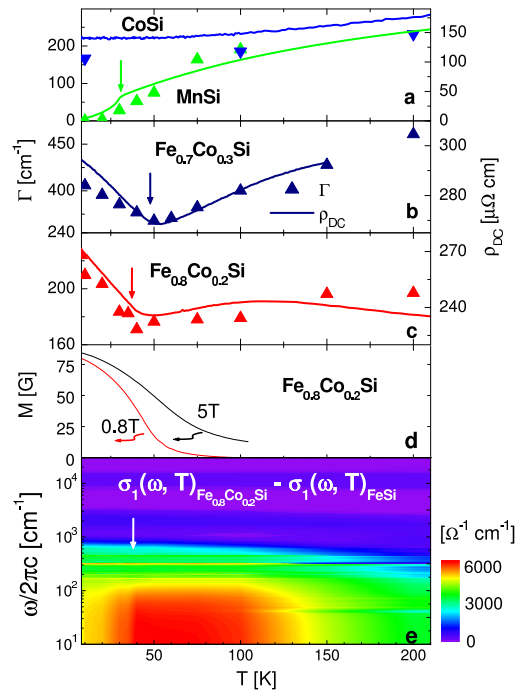}%
 \caption{\label{fig:tdependence} Low temperature scattering rate ($\Gamma$), 
DC resistivity ($\rho$),
magnetization ($M$), and optical conductivity ($\sigma(\omega)$).
(a), (b), and (c) $\rho$ (solid lines) and $\Gamma$ (triangles).
(d) $M$. (e) Difference between $\sigma(\omega)$ of
Fe$_{0.8}$Co$_{0.2}$Si and FeSi as function of $\omega$ and $T$.
Data and fit in Fig.\ref{fig:sigma} have been used. At
low-$\omega$ an additional decrease in $\sigma(\omega)$ can be
seen below $T_C$ in Fe$_{0.8}$Co$_{0.2}$Si. The arrows in parts a,
b, c, and e are $T_C$. }
 \end{figure}

Fig. \ref{fig:tdependence} and the insets in 
the middle frames of Figs.\ 2 and 3 
reveal more detail on the evolution of
the optical data with $T$, and allow comparison to transport and
magnetization results. The reflectivity (middle frame inset of Fig.\ 2) at low-$\omega$ 
simply follows the D.C. conductivity, which experiences its main drop 
below $T_C$. For higher $\omega$, the reflectivity decreases continuously 
from 300 K, with no visible anomaly at $T_C$. 
The disappearance of anomalies near $T_C$ as $\omega$ increases agrees with
the extended critical regime, or superparamagnetism (field induced
short range order), indicated by the magnetization data of
Fig. \ref{fig:tdependence}d. Here a modest (compared to $k_BT$)
external field of 5 T produces very appreciable polarization to
$T$'s as high as 100 K$\sim 3T_C$. Fig. \ref{fig:tdependence}e
shows $\sigma(\omega,T)$ of Fe$_{0.8}$Co$_{0.2}$Si after
subtraction of $\sigma(\omega)$ of FeSi, revealing where the added
carriers reside in the excitation spectrum of the nominally pure
compound. Cooling builds up the differential (relative to the
insulator) spectral weight down to $T_C$, whereupon there is a
loss especially apparent below 200 cm$^{-1}$. Again, our
observation of a reduced spectral weight below $T_C$ contradicts
both the standard models based on independent quasiparticles and
measurements for other magnetic semiconductors \cite{gamnasop}.

In addition, cooling below $T_C$ causes the wide Drude-like peak
to broaden, rather than to narrow.  We have parameterized the
low-$\omega$  $\sigma(\omega)$ by fitting the standard Drude form,
superposed on a $T$-independent Lorentzian peak at 800 cm$^{-1}$, to
the data. Figs.\ 4a-c display the resulting $\Gamma$'s, which
track the bulk resistivities ($\sigma_{DC}^{-1}$) and undergo a sharp
upturn below $T_C$. The important and unique contribution of the
optical data is to show that the unusual rise in resistivity below
$T_C$ is due as much to enhanced scattering as to a reduction in
the DOS.

To begin to understand our data, we have derived 
$\sigma(\omega)$ (Fig. \ref{fig:sigma}) from the band 
structure (see Fig. 1 and its caption) of
Fe$_{1-y}$Co$_y$Si. The band structure, computed assuming that 
Co doping only increases the effective charge at the iron site, 
agrees with the
magnetization and Hall effect and more complicated supercell
calculations, in that the minority spin Fermi surface resides in a
gap \cite{manyala1,guevera}. It also supports the idea that Co
doping merely adds electrons to rigid bands inherited from the
FeSi parent, even revealing, as does experiment, that CoSi (with
one extra electron/Fe) is a very low $n$ metal.  Another important
consequence is that in the polarized state, $E_F$ of the majority
spins appears at a DOS minimum, while in the unpolarized state,
the DOS at $E_F$  is high. This result is probably responsible for
the generally reduced metallicity of Fe$_{1-y}$Co$_y$Si as it is
cooled.

While band theory (Fig. \ref{fig:dos}) has some successes, it fails
in many significant regards, explaining neither the rapid filling
of the gap of FeSi with $T$, nor the apparent loss of carriers at
low-$\omega$, and increase---instead of the conventional
decrease---in $\Gamma$ below $T_C$ of Fe$_{1-y}$Co$_y$Si. To make
progress, the Coulomb interactions need to be considered. How
these underpin the moment formation at modest $T$, as well as the
rapid filling of the gap in the parent compound via the paradigm
of the Kondo insulator, is discussed elsewhere \cite{urasaki}. The
new optical effect in the doped material reported here most likely
follows from changes in the effective Coulomb coupling due to
quantum mechanical interference of diffusing charge
carriers\cite{altshuler}.  This feature of disordered metals,
hitherto seen only in transport and tunneling, but witnessed
optically for the first time in our experiments, induces
square-root singularities in the DOS at $E_F$\cite{altshuler}.
Spin-polarization, either from external magnetic fields or a
spontaneous magnetization, shifts the singularities with respect
to $E_F$ resulting in a reduction of
$\sigma(\omega)$\cite{altshuler}. Thus  $\sigma(\omega)$ and
$R(\omega)$ display singular behavior at low $T$ and $\omega$ just
as we observe. Furthermore, increased Coulomb scattering is
apparent in the continuous rise of $\Gamma$ as $T$ is lowered
below $T_C$.

We have shown that the optical properties of Fe$_{1-y}$Co$_y$Si
are very different from those for (GaMn)As, even though bulk
properties such as the off-diagonal conductivity are remarkably
similar\cite{manyala1}. Doping produces an optical response
throughout the gap of FeSi, implying that we are dealing not with
an impurity band – as in (Ga,Mn)As---but rather with carriers
donated to the conduction band of FeSi, which cause its gap to
collapse. Finally, we have discovered an optical reflectivity
which decreases rather increases upon entering the spin-polarized
state. The corresponding rise in the scattering rate, uniquely
visible in the optical data, demonstrates that the origin of this
effect is the Coulomb interaction between electrons in a
disordered system.

This work was supported by the US NSF, a
Wolfson-Royal Society Award, the Basic Technologies
program, the Swiss National Science
Foundation through the NCCR 'Materials with Novel Electronic
Properties', and the Netherlands---FOM/NWO. We thank
A. A. Menovsky and A.I. Poteryaev for assistance with
the crystal growth and LDA calculations.

 \end{document}